\documentclass[preprint,5p]{elsarticle}

\usepackage{lineno}
\usepackage{hyperref}
\usepackage{amsmath,amssymb}
\usepackage{color} 
\modulolinenumbers[1]

\usepackage{url}
\usepackage[vertfit]{breakurl}

\usepackage{ulem}

\journal{NIM A Short communication}

\begin{document}

\begin{frontmatter}

\title{Reduction of contaminants originating from primary beam \\ by improving the beam stoppers in GARIS-II}

\author[riken]{S. Kimura}\corref{mycorrespondingauthor}
\cortext[mycorrespondingauthor]{Corresponding author}
\ead{sota.kimura@riken.jp}

\author[riken]{D.~Kaji}
\author[jaea,riken]{Y.~Ito}
\author[kek]{H.~Miyatake}
\author[riken]{K.~Morimoto}
\author[kek]{P.~Schury}
\author[kek]{M.~Wada}

\address[riken]{Nishina Center for Accelerator Based Science, RIKEN, Wako, 351-0198, Japan}
\address[jaea]{Advanced Science Research Center, Japan Atomic Energy Agency, Tokai, Ibaraki 319-1195, Japan}
\address[kek]{Wako Nuclear Science Center (WNSC), Institute of Particle and Nuclear Studies (IPNS), \\ High Energy Accelerator Research Organization(KEK), Wako, 351-0198, Japan}

\begin{abstract}
Two independent beam stoppers have been developed for improving the beam separation of the gas-filled recoil ion separator GARIS-II. Performance of these supplemental beam stoppers was  investigated by using the $^{208}$Pb ($^{18}{\rm O},3{\rm n}$) $^{223}$Th reaction.  A160-fold enhancement of the signal-to-noise ratio at the GARIS-II focal plane was observed.
\end{abstract}


\end{frontmatter}


Gas-filled recoil separators are a powerful tool for  collecting superheavy nuclei (SHN) due to their capacity to cancel out the large velocity dispersions of fusion-evaporation reaction products; they have been implemented at several facilities \cite{Subotic2002,Gregorich2013,Gates2011,Leino1995}.  At the RIKEN Nishina Center for Accelerator-Based Science, two gas-filled recoil ion separators, GARIS \cite{Morita1992} and GARIS-II \cite{Kaji2013}, have been used in the study of production and decay properties of SHN  and the nuclear chemistry. 

In order to measure the masses of SHN, an experimental campaign called SHE-mass-I, wherein GARIS-II was coupled with a multi-reflection time-of-flight mass spectrograph (MRTOF-MS) \cite{Schury2014}, has been initiated. Precise masses of neutron-deficient isotopes from Bi ($Z=83$) to Ac ($Z=89$)  \cite{Schury2017,Rosenbusch2018} as well as isotopes from Es ($Z=99$) to No ($Z=102$) \cite{Ito2018} have already been measured successfully. 

SHE-mass-I requires the energetic beams from GARIS-II to be converted to low-energy beams before mass analyses  with the MRTOF-MS.  This is accomplished via a cryogenic gas-cell (GC) and an ion transport system that includes a multiple ion trap suite \cite{Schury2017}.  Due to space charge effects, the ion extraction efficiency of the GC would be reduced as the incoming beam rate increases \cite{Takamine2005}.  A wide range of fusion evaporation reactions, with the primary beam intensity exceeding 2~p$\mu$A in some cases, have been used to produce exotic isotopes for the SHE-mass-I project.   In several cases, unwanted particles with non-negligible intensities were observed to reach the GARIS-II focal plane, possibly be reducing the system efficiency and even posing a risk of damage to the GC.  Thus an enhancement in the suppression of unwanted particles was required. A study of background reduction with a beam stopper on a nearly symmetric reaction has been reported in \cite{Saren2011}. We report here on the development of such an improved suppression system and evaluation of its performance with an asymmetric reaction.

\begin{figure}[!b]
  \begin{center}
    \includegraphics[width=0.4\textwidth, bb=0 0 842 595, clip, trim=90 80 90 80]{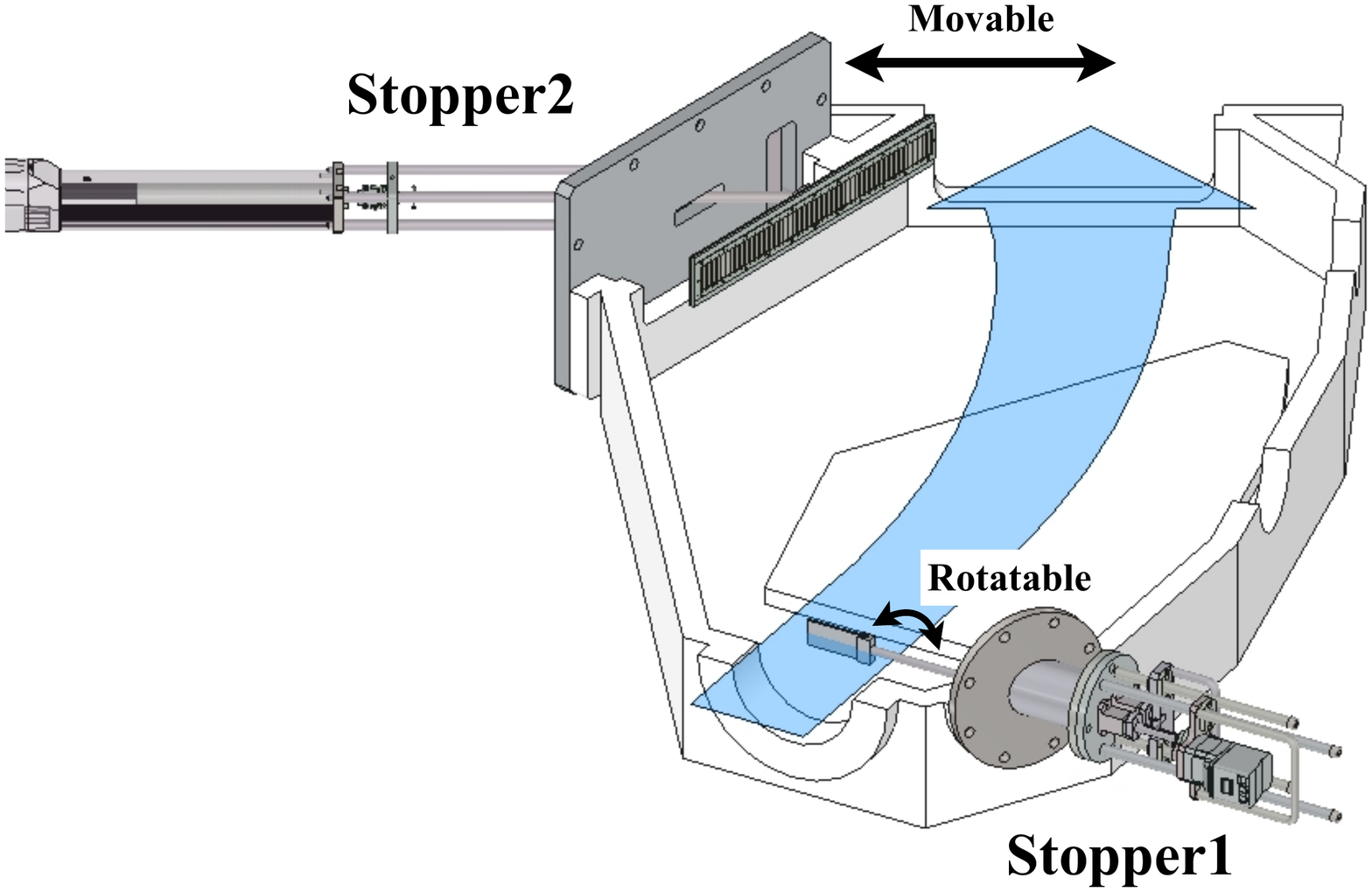}
  \end{center}
  \vspace*{-10pt}
  \caption{Schematic view of the primary beam stoppers. The D1 chamber is shown in cross-section. The blue arrow indicates the beam direction of evaporation residues. This figure presents an operational state where Stopper1 has an effective area of $24 \ {\rm cm}^2$ while Stopper2 is at the $0 \ {\rm cm}$ position.\label{SchematicView}}
\end{figure}

The primary beam and the evaporation residues (EVRs) differ in terms of their behavior in GARIS-II based on their $B\rho$-values and angular distributions immediately after recoiling out the target. Even though their $B\rho$-values are generally similar to each other, as discussed above, although there remains a capability to separate them by the small differences of their trajectories at the exit of the first dipole (D1) of GARIS-II. EVRs have large angular distributions compared to the primary beam, due mainly to multiple scatterings in the target materials and recoils of evaporated light-particles in de-excitation processes. As such, primary beam and EVRs could be separated by utilizing the differences in their angular distributions.

For separating the primary beams and the EvRs, two water-cooled primary beam stoppers, ``Stopper1" and ``Stopper2", were designed and installed in the D1 chamber of GARIS-II as shown in Fig.~\ref{SchematicView}.  Stopper1, installed at the entrance of the D1 chamber, consisted of a copper plate of $8 \ {\rm cm}$ width and $3 \ {\rm cm}$ height. It was possible to vary the effective area facing  to the beam by rotating the structure.  Stopper2, located near the D1 chamber exit, was made of a $40 \ {\rm cm}$ wide and $6 \ {\rm cm}$ high copper board mounted on a linear manipulator to change its position; tantalum fins on its surface to prevent scattering of impinging particles. 

The performance of the stoppers were evaluated by using the $^{208}$Pb ($^{18}{\rm O},3{\rm n}$) $^{223}$Th reaction. A 4.84~MeV/u $^{18}$O$^{5+}$ beam with average intensity of 16~pnA was provided by the RIKEN linear accelerator RILAC. The $^{208}$Pb target segments were prepared by evaporation onto 60~$\mu$g/cm$^2$ carbon backings until an average thickness of 450~$\mu$g/cm$^2$ was achieved.  The target segments were mounted on a 30 cm diameter target wheel \citep{Kaji2015} that was nominally rotated at 2000 rpm during irradiation. 

\begin{figure}[!t]
  \centering
     \includegraphics[width=0.4\textwidth, bb=0 0 842 595, clip, trim=180 20 185 25]{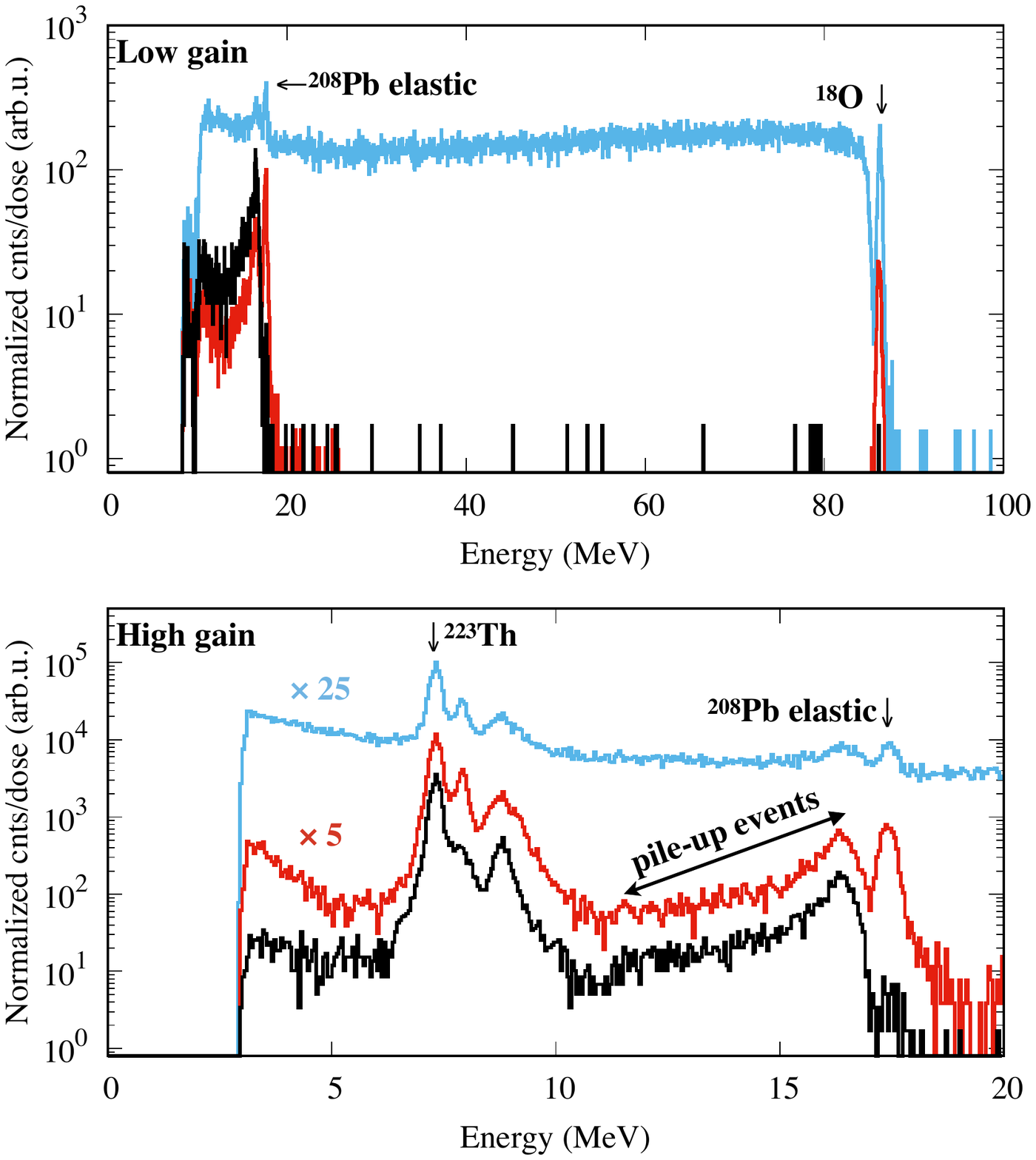}
  \caption{Energy spectra measured by the silicon detector array, (Blue line) with neither both stoppers nor primary beam chopping, (Black line) with primary beam chopping and without both stoppers,  and (Red line) with both stoppers  but without primary beam chopping, respectively. \label{Spectra}}
\end{figure}

An array of silicon detectors (HAMAMATSU S3204-09), arranged in a $3\times5$ pattern, was employed to count both the incoming $^{223}{\rm Th}$ and contaminants reaching the GARIS-II focal plane. Signals from  the detector array were processed with both high- and low-gain circuits. The primary beam was chopped to measure the $\alpha$-decay from $^{223}$Th under low-background conditions. The beam chopping sequence was chosen to be 0.2~s beam-on and 0.2~s beam-off.  Energy spectra measured by both high- and low-gain circuits are shows in Fig~\ref{Spectra}.  The high-gained data were used for counting the $^{223}$Th events, where the intensities were evaluated with Gaussian distribution . The intensities of the contaminants were obtained from integration of the low-gain data down to 17~MeV, which is a border between the $^{208}$Pb elastic events and the pile-up events of fast decay of $^{223}{\rm Th}$'s granddaughter $^{215}{\rm Rn}$ ($t_{1/2} = 2.3~\mu{\rm s}$). Because of this ambiguity, low energy contaminant events are not included in the present analysis. Counts of $^{18}$O beam and $^{208}$Pb elastic events investigated with Gaussian distributions for the low- and high-gained data, respectively. 

\begin{figure}[!t]
  \centering
    \includegraphics[width=0.5\textwidth, bb=0 0 842 595, clip, trim=15 120 30 120]{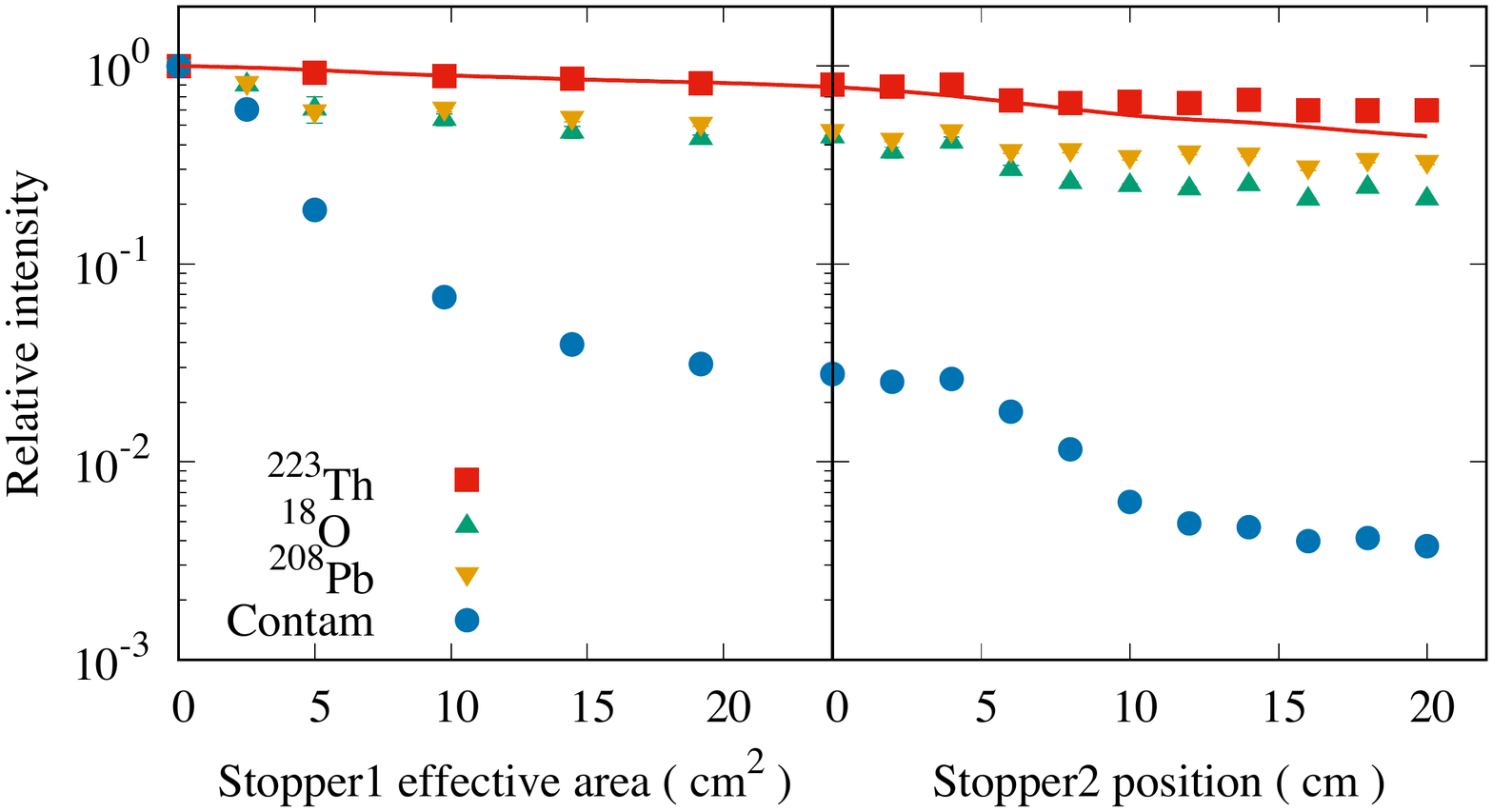}
  \caption{Dependencies of the counting rates on the Stopper1 effective area (left) and on the Stopper2 position (right). Size of the all error bars are smaller than the points. Red line represents the result of $^{223}$Th transport simulations. \label{Results}}
%
  \centering
    \includegraphics[width=0.45\textwidth, bb=0 0 842 595, clip, trim=0 0 0 0]{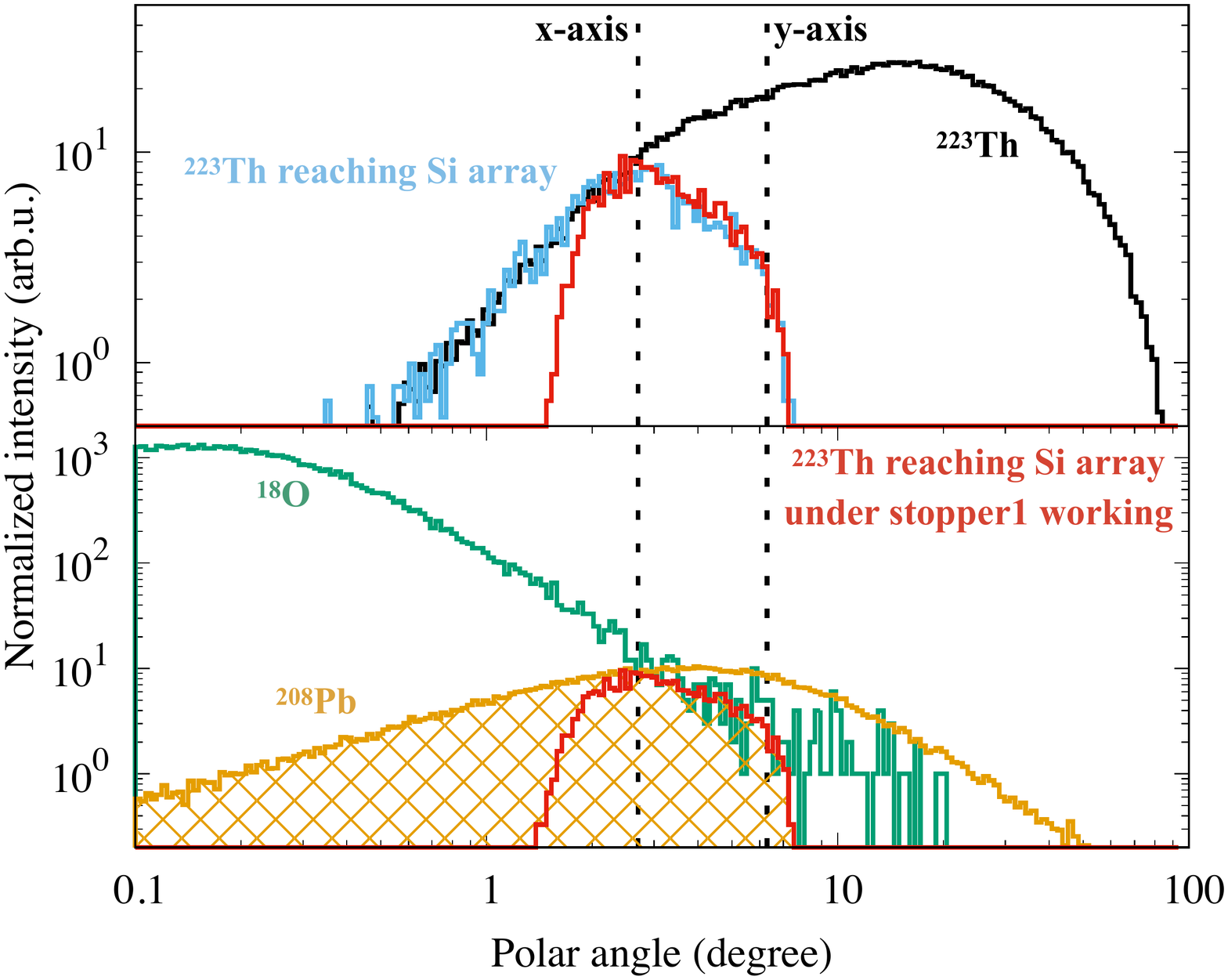}
  \caption{Simulated polar angle distributions of emitted particles right after the target. (Upper panel, black line) Simulated result of $^{223}$Th without gate. (Upper panel, blue line) $^{223}$Th result gated with a condition that reaching the silicon detector array installed at the GARIS-II focal plane. (Upper and lower panels, red lines) Same gate as the blue line, but simulated result under the assumption that stopper1 fully works. (Lower panel, green line) Simulated result of $^{18}$O. (Lower panel, orange line) Result of $^{208}$Pb assuming zero-degree recoil. (Dotted lines) GARIS-II acceptance of x- and y-axis. See the text for the details of simulations. \label{AngleDist}}
\end{figure}

Due to relatively long mean-free-path as longer than 10~m, sizable fraction of $^{18}{\rm O}^{4+}$ ion could be survived through the flight path of the D1 chamber and directly transported to the GARIS-II focal plane,  since the $B\rho$-value is close to the $^{223}$Th: $B\rho \left( ^{18}{\rm O} (q=+4) \right) / B\rho \left( ^{223}{\rm Th}(\bar{q}=+3.4) \right) = 0.985$. Besides this, a multi-scattering process with the filling gas could be considered as the reason of direct transportation of primary beam. An intensity ratio of the contaminants to the $^{223}$Th was around 20 without either stopper.  As the Stopper1 effective area began to increase, the intensity of the contaminants rapidly decreased; in contrast, the $^{223}{\rm Th}$ counting rate varied only gradually (left panel of Fig.~\ref{Results}). As a result, we could suppress $\sim 97\%$ of the contaminants by using Stopper1, while $\sim 80\%$ of $^{223}{\rm Th}$ was successfully transported. The same measurements were conducted by changing the Stopper2 position as shown in the right panel of Fig.~\ref{Results}. When measuring the Stopper2 performance, the Stopper1 effective area was set to its maximum value of $24 \ {\rm cm}^2$.  The Stopper2 measurement shows results similar to those obtained with Stopper1. Thus, in the end, we can improve the signal-to-noise ratio by a factor of $\sim160$ by using both stoppers. 

\begin{figure}[!t]
  \centering
    \includegraphics[width=0.45\textwidth, bb=0 0 360 252, clip, trim=0 30 0 25]{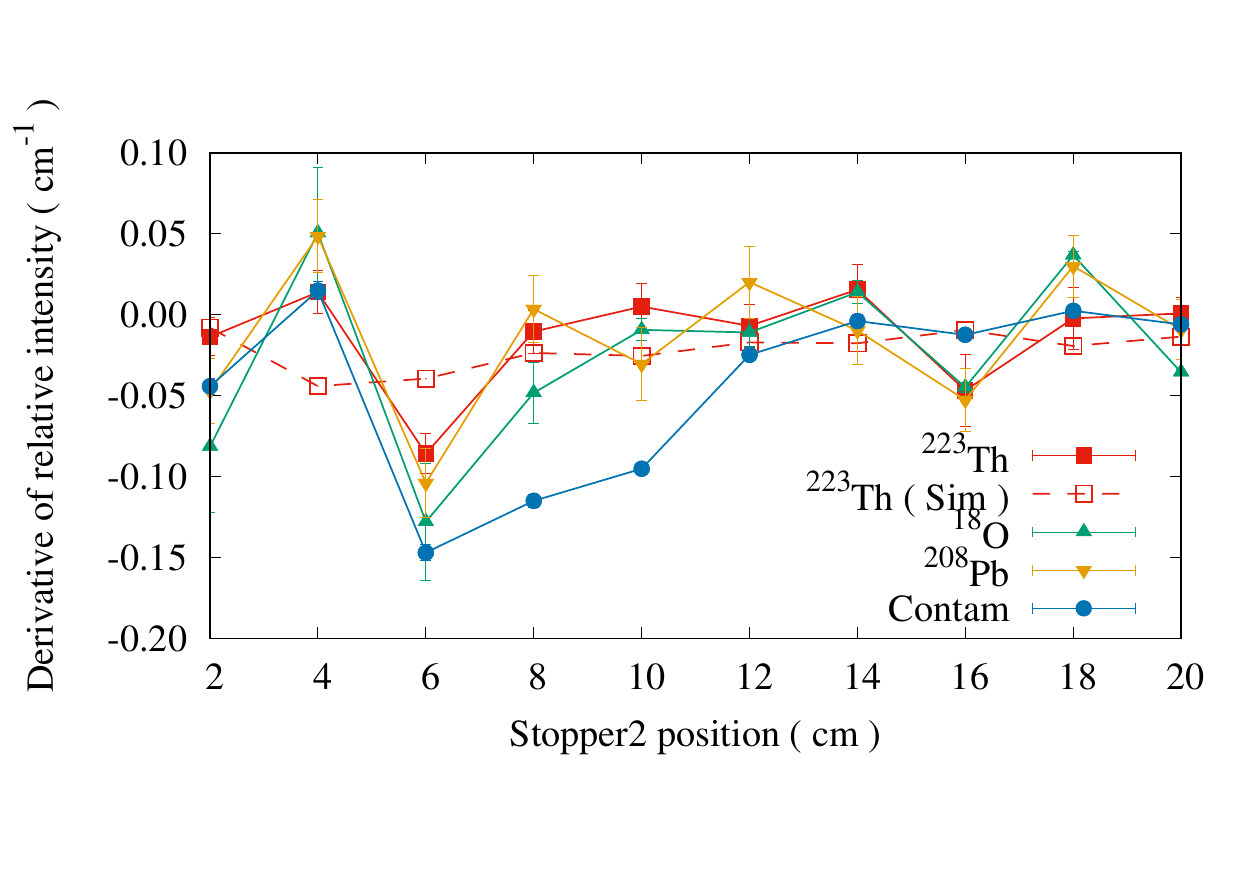}
  \caption{Derivatives of the relative intensities (as shown in Fig.~\ref{Results}) respecting to the Stopper2 position. To calculate the derivative, the relative intensity was set to 1 at Stopper2 position equals 0~cm. Red open box and dashed line show the simulated result of $^{223}$Th. \label{Stopper2_deriv}}
\end{figure}

Red line shown in Fig.~\ref{Results} indicates the simulated result of $^{223}$Th transport in GARIS-II based on Geant4 \cite{Agostinelli2003}, including consideration of recoils of evaporated particles in the de-excitation process, charge exchange processes between the GARIS-II buffer gas and ions, scatterings by materials including the buffer gas, and the GARIS-II geometrical conditions including both the stoppers and other relevant devices. The simulation reproduces the trend of reduction of $^{223}$Th transmission efficiency by the use of Stopper1. Figure~\ref{AngleDist} shows the simulated polar angular distributions  of emitted particles right after the target. This indicates that $^{223}$Th events only having limited scattering angle can reach the GARIS-II focal plane. The reduction rate of $^{208}$Pb elastic events by Stopper1 can be explained via Fig.~\ref{AngleDist}. Assuming that the $^{208}$Pb events reaching the GARIS-II focal plane have the same angular distribution as the $^{223}$Th events, the reduction rate can be represented by a ratio of the area surrounded by the red line to the orange-colored mesh area and is found to be $45\%$. This value is consistent  with the experimental value of $47.3(2.4)\%$. But, in contrast with this, the reduction rate of $^{18}$O cannot be explained in the same way. The reason for this is incompletely understood and further investigation is necessary. 

Figure.~\ref{Stopper2_deriv} shows the reduction trend of each component as changing the Stopper2 position to increase its effect. Their behavior should be affected to position distribution near the D1 exit. Non uniform reduction trends indicate that the position distributions of relevant components would have elliptical structures which reflect ion loss by the Stopper1. As indicating Fig.~\ref{Stopper2_deriv}, the continuous component has the wider position distribution comparing with others. This can be understood as follows: assuming that the continuous component consists of beam-like particles and have significant long mean-free-path of charge exchange reaction to ignore the charge state equilibration, their trajectories spread in the D1 magnet due to the wide kinetic energy distribution. For $^{223}$Th, its position distribution is localized because of its charge state reaches equilibrium state and it has narrow $B\rho$-value width. Same argument would valid for the case of $^{208}$Pb. As same as the continuous component,  $^{18}$O beam has long mean-free-path of charge exchange reaction but has the certain kinetic energy. Then $^{18}$O beam component has localized position distribution at the D1 exit. 

The contaminant events can be sorted into two categories: peaks and continuous as indicated in Fig~\ref{Spectra}. Both stoppers mainly reduce the continuous component. For the production mechanisms of the continuous component having wide energy range, some mechanisms such as multiple elastic and quasi-elastic scatterings with the target and any materials, $i.e.$ vacuum chamber inner wall, can be considered. But it is difficult to identify the origin of the continuous component based the information of energy spectra only. Thus further investigations are necessary to understand the origin of the contaminants and achieve a maximal low background environment at the GARIS-II focal plane. 

In order to improve the signal-to-noise ratio at the GARIS-II focal plane, two independent beam stoppers have been developed. Performance evaluations of the beam stoppers were performed by using the $^{208}$Pb ($^{18}{\rm O},3{\rm n}$) $^{223}$Th reaction. This study indicates a 160-fold enhancement of the signal-to-noise ratio at the GARIS-II focal plane has been achieved for the asymmetric reaction, $^{208}$Pb ($^{18}{\rm O},3{\rm n}$) $^{223}$Th, by using the beam stoppers. We demonstrate that a simple mechanism can drastically reduce the contaminants at the GARIS-II focal plane while maintaining the intensity of the desired fusion-evaporation reaction products. An initial application of the beam stoppers is reported in \cite{Kimura2018}. \\

We would like to express our sincere gratitude to the RIKEN Nishina Center for Accelerator-Based Science and the Center for Nuclear Science at the University of Tokyo for their support of the present measurements. This study was supported by the Japan Society for the Promotion of Science KAKENHI, Grant Number 24224008, 15H02096, 15K05116, and 17H06090.

\bibliography{beam_stopper}

\end{document}